\documentclass[twocolumn,aps,prd,showpacs,superscriptaddress]{revtex4}
\usepackage{amsmath}
\usepackage{graphicx}
\usepackage[usenames]{color}
\usepackage[letterpaper,margin=0.75in]{geometry}
\usepackage{multirow}

\usepackage[colorlinks,bookmarks]{hyperref}
\definecolor{linkblue}{rgb}{0,0,0.8}
\definecolor{linkgreen}{rgb}{0,0.5,0}
\hypersetup{pdfpagemode=None, pdfstartview=FitH, linkcolor=linkblue, %
             citecolor=linkgreen, urlcolor=linkblue}
\newcommand{\sorthelp}[1]{}

\def\smica{{\tt SMICA}}
\def\beq{\begin{equation}}
\def\eeq{\end{equation}}
\def\bea{\setlength\arraycolsep{1.4pt}\begin{eqnarray}}
\def\eea{\end{eqnarray}}
\def\bit{\begin{itemize}}
\def\eit{\end{itemize}}

\def\eq{Eq.~}

\def\fig{Fig.~}

\def\ld{\left}
\def\rd{\right}

\def\wtl{\widetilde}

\def\fr{\frac}

\def\sig{\sigma}

\def\L{{\cal L}}

\def\bra{\langle}
\def\ket{\rangle}

\def\LCDM{$\Lambda$CDM}
\def\Planck{\textit{Planck}}

\def\dhat{{\hat{\mathbf{d}}}}

\def\nhat{{\hat{\mathbf{n}}}}
\def\vk{{\boldsymbol{k}}}
\def\vx{{\boldsymbol{x}}}
\def\Ph{{{\cal P}^{\rm hi}}}
\def\Pl{{{\cal P}^{\rm lo}}}

\newcommand{\pz}{\phantom{0}}

\begin{document}

\title{Closing in on the large-scale CMB power asymmetry}

\author{D. Contreras} \email{dagocont@phas.ubc.ca}
\affiliation{Department of Physics \& Astronomy\\
University of British Columbia, Vancouver, BC, V6T 1Z1  Canada}

\author{J. Hutchinson} \email{jhutchin@ualberta.ca}
\affiliation{Department of Physics\\
University of Alberta, Edmonton, AB, T6G 2E1  Canada}

\author{A. Moss} \email{adam.moss@nottingham.ac.uk}
\affiliation{Centre for Astronomy \& Particle Theory\\
University of Nottingham, University Park, Nottingham, NG7 2RD  U.K.}

\author{D. Scott} \email{dscott@phas.ubc.ca}
\affiliation{Department of Physics \& Astronomy\\
University of British Columbia, Vancouver, BC, V6T 1Z1  Canada}

\author{J. P. Zibin} \email{zibin@phas.ubc.ca}
\affiliation{Department of Physics \& Astronomy\\
University of British Columbia, Vancouver, BC, V6T 1Z1  Canada}

\date{\today}

\begin{abstract}

   Measurements of the cosmic microwave background (CMB) temperature
anisotropies have revealed a dipolar asymmetry in power at the largest scales,
in apparent contradiction with the statistical isotropy of standard
cosmological models.  The significance of the effect is not very high, and
is dependent on {\em a posteriori} choices.  Nevertheless, a number of models
have been proposed that produce a scale-dependent asymmetry.  We confront
several such models for a physical, position-space modulation with CMB
temperature observations.  We find that, while some models that maintain the
standard isotropic power spectrum are allowed, others, such as those with
modulated tensor or uncorrelated isocurvature modes, can be ruled out on the
basis of the overproduction of isotropic power.  This remains the case even
when an extra isocurvature mode fully anti-correlated with the adiabatic
perturbations is added to suppress power on large scales.
\end{abstract}

\maketitle

%----------------- INTRODUCTION -----------------------
\section{Introduction}

%{\em Introduction.---}%
   The standard six-parameter $\Lambda$ cold dark matter (\LCDM) cosmological
model describes the temperature fluctuations in the cosmic microwave
background (CMB) radiation spectacularly well, as demonstrated by the
WMAP satellite~\cite{wmap9}, the Atacama Cosmology Telescope~\cite{act12},
the South Pole Telescope~\cite{spt12}, and, especially, the \Planck\
satellite~\cite{planck2014-a01}.  Central assumptions in the \LCDM\ model are
that the
fluctuations are Gaussian and statistically homogeneous and isotropic.
Despite the success of the standard model, several ``anomalies'' have been
noticed in the CMB, which apparently violate these assumptions (for reviews,
see Refs.~\cite{Bennett2011,planck2013-p09,planck2014-a18,Schwarzetal2016}).  The
statistical significance of these anomalies is not very high, and is weakened
substantially with {\em a posteriori} (look elsewhere)
corrections~\cite{zh10,Bennett2011,bunn10,planck2014-a18} when those are well
defined.

   Probably the most intriguing of the anomalies is a very roughly $6\%$
dipolar or hemispherical asymmetry in the large-scale CMB temperature
fluctuation power,
first noted in the WMAP one-year data~\cite{Eriksen2004}.  Later analyses
showed that the asymmetry is substantially reduced on multipole scales $\ell
\gtrsim 100$~\cite{hansenetal09,Hanson2009,fh13,planck2014-a18,qn15,awkkf15}.
The significance of the asymmetry is only at the $3\sig$ level,
and is sensitive to {\em a posteriori} choices in the maximum multipole
scale~\cite{Bennett2011,planck2014-a18}, so
it should perhaps not be considered a great surprise.  Nevertheless, an origin
to the asymmetry as a physical modulation of the primordial fluctuations
would clearly be of fundamental importance for cosmology, and in particular
might provide information about inflation, given the large-scale nature of
the effect.  Therefore it is worthwhile to investigate possible physical
explanations.  Since concrete inflationary models for modulation are
difficult to construct~\cite{brst16}, we consider phenomenological models in
this study.

   While most studies of the dipolar asymmetry have been performed in
angular multipole space, any physical model will necessarily be described
best in position (or $k$) space.  In Ref.~\cite{Zibin2015} we developed a
formalism for describing a spatial modulation and its effect on CMB temperature
anisotropies, and for performing Bayesian estimation of the modulation
parameters.  This formalism was crucial for answering an important question:
what does a modulation that fits the temperature data predict for
{\em other} observations, such as CMB lensing~\cite{Zibin2015} and
polarization~\cite{long}?  Given the inconclusive significance level of the
asymmetry, probes of modes independent from CMB temperature may be essential
in order to confirm or refute a physical origin to the asymmetry.  Our
formalism is an extension of an approach to describe the effects in the CMB
of gradients in cosmological parameters~\cite{Moss2011}.  The effects of
various such parameter gradients were discussed in~\cite{pesky}.

   In this paper we apply our formalism to determine whether any models for
modulation can already be ruled out on the basis of current CMB temperature
data.  We point out that some models necessarily increase the statistically
{\em isotropic} temperature power, and so the ordinary power spectra can
be considered as ``independent probes'' to test a physical origin for the
asymmetry.
We consider purely phenomenological modulations of the ordinary
adiabatic fluctuations, as well as a gradient of the scalar spectrum tilt
and modulations of tensor and isocurvature contributions, in doing so
testing several of the models discussed in~\cite{pesky}.  Modulated
isocurvature modes were studied in~\cite{Hanson2009}, and, in the context
of a particular inflationary model~\cite{ehk09}, in Ref.~\cite{planck2014-a24}.
In the process we also provide constraints on {\em un}modulated tilted
tensor and isocurvature modes using the latest data.

%-----------------------GENERAL COVARIANCE--------------------------------
\section{Formalism}
%\label{sec:form}

%{\em Formalism.---}%
   Our goal is to construct physical, position-space models for a
temperature dipolar asymmetry, which is confined mostly to large scales.
We apply the formalism developed in
Refs.~\cite{Zibin2015,long}, which captures scale dependence by employing two
fluctuation components.  The first, $\wtl Q^{\rm lo}(\vx)$, is restricted
mainly to large scales (low $k$) and is maximally linearly spatially modulated:
\beq
\wtl Q^{\rm lo}(\vx) = Q^{\rm lo}(\vx)\ld(1 + \fr{\vx\cdot\dhat}{r_{\rm LS}}\rd),
\label{modflucts}
\eeq
where $Q^{\rm lo}(\vx)$ is statistically isotropic with power spectrum
$\Pl(k)$, $\dhat$ is the direction of modulation, and $r_{\rm LS}$ is the
comoving distance to last scattering.  The total modulation amplitude will be set
by a multiplicative factor inside $\Pl(k)$ \footnote{This convention differs
trivially from that used in~\cite{Zibin2015,long}, in which the parameter $A$
is equivalent to our parameters $A_{\tanh}$ or $A_{\rm PL}$.}.  The second
component, $Q^{\rm hi}(\vx)$, is statistically isotropic with spectrum
$\Ph(k)$.  The two fields are taken to be uncorrelated, i.e.,
$\bra Q^{\rm lo}(\vk)Q^{\rm hi*}(\vk')\ket = 0$.  We attempt to be agnostic
as to the origin of the modulation;
the isotropic $Q^{\rm hi}$ component is
adiabatic, while for the modulated component, $Q^{\rm lo}$, we consider
adiabatic, CDM isocurvature, and tensor fluctuations.

   The total temperature anisotropies due to these two fields will
be to a very good approximation~\cite{Zibin2015}
\beq
\delta T(\nhat) = \delta T^{\rm lo}(\nhat)\ld(1 + \nhat\cdot\dhat\rd)
                + \delta T^{\rm hi}(\nhat),
\label{eq:totTanis}
\eeq
where $\delta T^{\rm lo}$, with power spectrum $C_\ell^{\rm lo}$
(called the ``asymmetry spectrum''), is produced by $\Pl(k)$, while
$\delta T^{\rm hi}$, with spectrum $C_\ell^{\rm hi}$, is produced by
$\Ph(k)$.  These anisotropies lead to the lowest-order spherical harmonic
multipole covariance~\cite{pubb05,Moss2011,planck2014-a18,Zibin2015}
\beq
\bra a_{\ell m}a_{\ell'm'}^*\ket = C_\ell\delta_{\ell\ell'}\delta_{mm'}
   + \fr{\delta C_{\ell\ell'}}{2}\sum_M d_M\xi^M_{\ell m\ell'm'},
\label{eq:cmbcovariance}
\eeq
where $\delta C_{\ell\ell'} \equiv 2(C_\ell^{\rm lo} + C_{\ell'}^{\rm lo})$
and $d_M$ is the spherical harmonic decomposition of $\nhat\cdot\dhat$.
The coefficients $\xi^M_{\ell m\ell'm'}$ couple modes $\ell$ to $\ell \pm 1$:
\beq
\xi^M_{\ell m\ell'm'} \equiv \sqrt{\fr{4\pi}{3}}
   \int Y_{\ell'm'}(\nhat)Y_{1M}(\nhat)Y_{\ell m}^*(\nhat)d\Omega.
\eeq

   Crucially, the modulated component will also contribute to the total
isotropic power, via
\beq
C_\ell = C_\ell^{\rm lo} + C_\ell^{\rm hi}.
\eeq
Therefore a model that produces sufficient asymmetry to fit the temperature
data may overproduce isotropic power at large scales and hence be
inconsistent with experiments such as \Planck.

\section{Models}

%{\em Models.---}%
   We employ the same models as described in Ref.~\cite{long} to describe a
large-scale modulation.  First, we consider the adiabatic $\tanh$ model,
with $k$-space asymmetry spectrum
\beq
  \mathcal{P}^{\rm lo}(k)%_{\mathcal{R}}
  = \frac{A_{\tanh}}{2} \mathcal{P}^{\Lambda{\rm CDM}}(k)%_\mathcal{R}
     \left[1 - \tanh{\left(\frac{\ln k - \ln k_{\rm c}}{\Delta\ln k}\right)}\right],
\eeq
where
\beq
\mathcal{P}^{\Lambda{\rm CDM}}(k)%_\mathcal{R}
   = A_{\rm s}\left(\frac{k}{k_0}\right)^{n_{\rm s} - 1}
\label{LCDMPS}
\eeq
describes the usual \LCDM\ power-law primoridal comoving curvature
perturbation spectrum.  The parameters $\Delta\ln k$ and $k_{\rm c}$ describe
the width and position of a small-scale cutoff and $A_{\tanh} \le 1$ is
the amplitude of the modulation.  Next we consider an adiabatic
power-law model (abbreviated ``ad.-PL''):
\beq
\mathcal{P}^{\rm lo}(k)%_\mathcal{R}
   = A_{\rm PL}\mathcal{P}^{\Lambda{\rm CDM}}(k_0^{\rm lo})%_\mathcal{R}
     \left(\frac{k}{k_0^{\rm lo}}\right)^{n^{\rm lo}_{\rm s} - 1},
   \label{eq:powerlaw}
\eeq
where $n^{\rm lo}_{\rm s}$ and $A_{\rm PL} \le 1$ are the modulation tilt and
amplitude, and $k_0^{\rm lo} = 1.5\times10^{-4}\,{\rm Mpc}^{-1}$ is a pivot
scale.  For both of these adiabatic models we fix $\Ph(k)$ via the constraint
\beq
\mathcal{P}^{\rm lo}(k) + \mathcal{P}^{\rm hi}(k)
   = \mathcal{P}^{\Lambda{\rm CDM}}(k)
\label{lohiLCDM}
\eeq
(and hence $C_\ell^{\rm lo} + C_\ell^{\rm hi} = C_\ell^{\Lambda{\rm CDM}}$),
so that the isotropic power is automatically consistent with standard \LCDM.

   Next we consider a single-component adiabatic model with a linear gradient
in the tilt, $n_{\rm s}$, of the primordial power spectrum (``$n_{\rm s}$-grad''
for short).  In this case we can directly write the asymmetry spectrum
as~\cite{Zibin2015,long}
\beq
C_\ell^{\rm lo} = -\fr{\Delta n_{\rm s}}{2}\fr{dC_\ell^{\Lambda{\rm CDM}}}{dn_{\rm s}}.
\eeq
Here we have used a linear approximation for the effect of the gradient, which
will be well justified by our results.  The modulation amplitude is specified
by the increment in tilt, $\Delta n_{\rm s}$, from modulation equator to
pole.  Note that this modulation will depend implicitly on the pivot scale
for $A_{\rm s}$.

   Finally we consider three models that naturally produce contributions on
large scales.  The first is a modulation of the standard $\Lambda$CDM
integrated Sachs-Wolfe (ISW) contibution
with amplitude $A_{\rm ISW} \leq 1$ \footnote{Note that the ISW effect is
sourced over a wide range of distances, so it is unlikely that a
position-space modulation could result in maximal ISW modulation, i.e.,
$A_{\rm ISW} = 1$.  Therefore
our results will be conservative, in that a realistic ISW modulation would
likely produce less asymmetry.}.  This phenomenological model automatically
satisfies \emph{isotropic} CMB constraints and $C^{\rm lo}_\ell$ is simply the
contribution of the ISW effect to the total power $C_\ell$.
The second is a modulated CDM density isocurvature component,
\beq
\Pl(k)% = \mathcal{P}_\mathcal{I}(k)
   = \frac{\alpha_{k_*}}{1 - \alpha_{k_*}}
     \mathcal{P}^{\Lambda{\rm CDM}}(k_*)
     \left(\frac{k}{k_*}\right)^{n_{\mathcal{I}} - 1},
\label{eq:isomodel}
\eeq
and the third is a modulated tensor component,
\beq
\Pl(k)% = \mathcal{P}_{\rm t}(k)
   = r_{k_*}\mathcal{P}^{\Lambda\rm{CDM}}(k_*)
     \left(\frac{k}{k_*}\right)^{n_{\rm t}}.
\label{eq:tensormodel}
\eeq
In these latter two cases the models are described by two parameters, a primordial
power ratio ($\alpha_{k_*}$ or $r_{k_*}$, evaluated at scale $k_* = 0.002
\,\mathrm{Mpc}^{-1}$) and a
tilt ($n_{\mathcal{I}}$ or $n_{\rm t}$).  For both isocurvature and tensor
models we fix $\Ph(k) = \mathcal{P}^{\Lambda{\rm CDM}}(k)$, so that the
additional isotropic power from the modulated component will further
constrain these models.  For the tensor model we also consider an
unmodulated isocurvature component that is fully \mbox{(anti-)correlated}
with the adiabatic scalars.  Anti-correlated isocurvature modes would
decrease power on large scales, potentially allowing for a larger
contribution of modulated tensors.  This inclusion adds one
extra parameter, which is simply the amplitude of perturbations for the new mode.

%-----------------GRADIENT ESTIMATOR------------------------------------------
\section{Modulation estimator}
%\label{sec:modest}

%{\em Modulation estimator.---}%
   For a full-sky, noise-free measurement of the temperature multipoles, we
can write down an estimator for the modulation amplitude
$\Delta X_M \equiv Ad_M$ as~\cite{Moss2011,planck2014-a18,Zibin2015}
\beq
\Delta\hat{X}_M = \fr{1}{4A}\sigma_X^2 \sum_{\ell m \ell' m'} \fr{\delta
C_{\ell\ell'}}{C_\ell C_{\ell'}} \xi^M_{\ell m\ell'm'}a_{\ell m}^*a_{\ell'm'},
\eeq
where $A = A_{\tanh}$, $A_{\rm PL}$, $\Delta n_{\rm s}$, $A_{\rm ISW}$,
$\alpha_{k_*}/(1 - \alpha_{k_*})$, or $r_{k_*}$, depending on the model, and
where the cosmic variance of the estimator is given by
\beq
\sigma_X^2 = 12A^2\ld(\sum_\ell(\ell + 1)
             \fr{\delta C_{\ell\ell + 1}^2}{C_\ell C_{\ell + 1}}\rd)^{-1}.
\label{eq:sigmax}
\eeq
The presence of noise and incomplete sky coverage modifies the above relations.
We use a $C$-inverse filter approach that accounts for noise, and, optimally,
for the mask (as described in Refs.~\cite{planck2013-p12,
planck2014-a17}). Masking and residuals in the data will induce a mean-field
value for $\Delta X_M$ that can be estimated with simulations. Further details
of the full estimator we use can be found in Appendix C of
Ref.~\cite{planck2014-a18}.

For fixed modulation parameters the maximum likelihood is
\beq
\ln\L = \sum_M\fr{\Delta\hat{X}_M^2}{2\sig_X^2}.
\label{lnLanis}
\eeq
We can then build the rest of the likelihood by sampling on a grid of values
for the $k$-space parameters (see Ref.~\cite{Zibin2015}). For the tensor and
isocurvature models we assign a uniform prior on $A$, in order to obtain
consistency with the \emph{isotropic} likelihood results. For all other models
we use a prior uniform in the individual $\Delta X_M$.

%----------------Results-----------------------
\section{Results}
%\label{sec:results}

\begin{figure}[h]
\centerline{\includegraphics[width=\hsize]{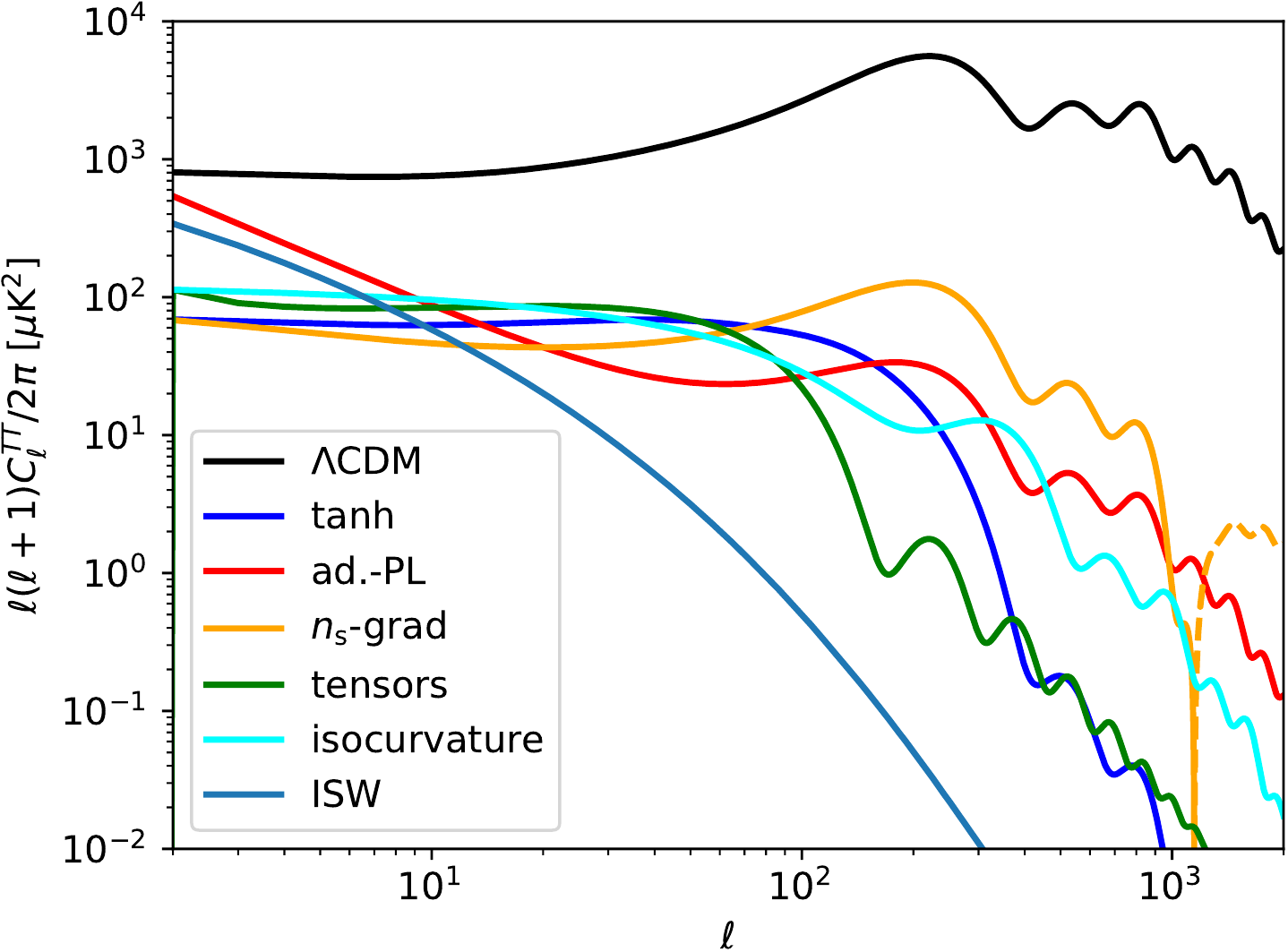}}
\caption{\LCDM\ temperature spectrum compared to the best-fit asymmetry
spectra, $C_\ell^{\rm lo}$, for the various models.  The best fits correspond
roughly to a 5--10\% asymmetry for $\ell \lesssim 100$, as expected, with the
exception of the ISW modulation, whose maximum amplitude (and shape) is
fixed by \LCDM.}
\label{fig:bestfits}
\end{figure}

\begin{figure}[h]
\centerline{\includegraphics[width=\hsize]{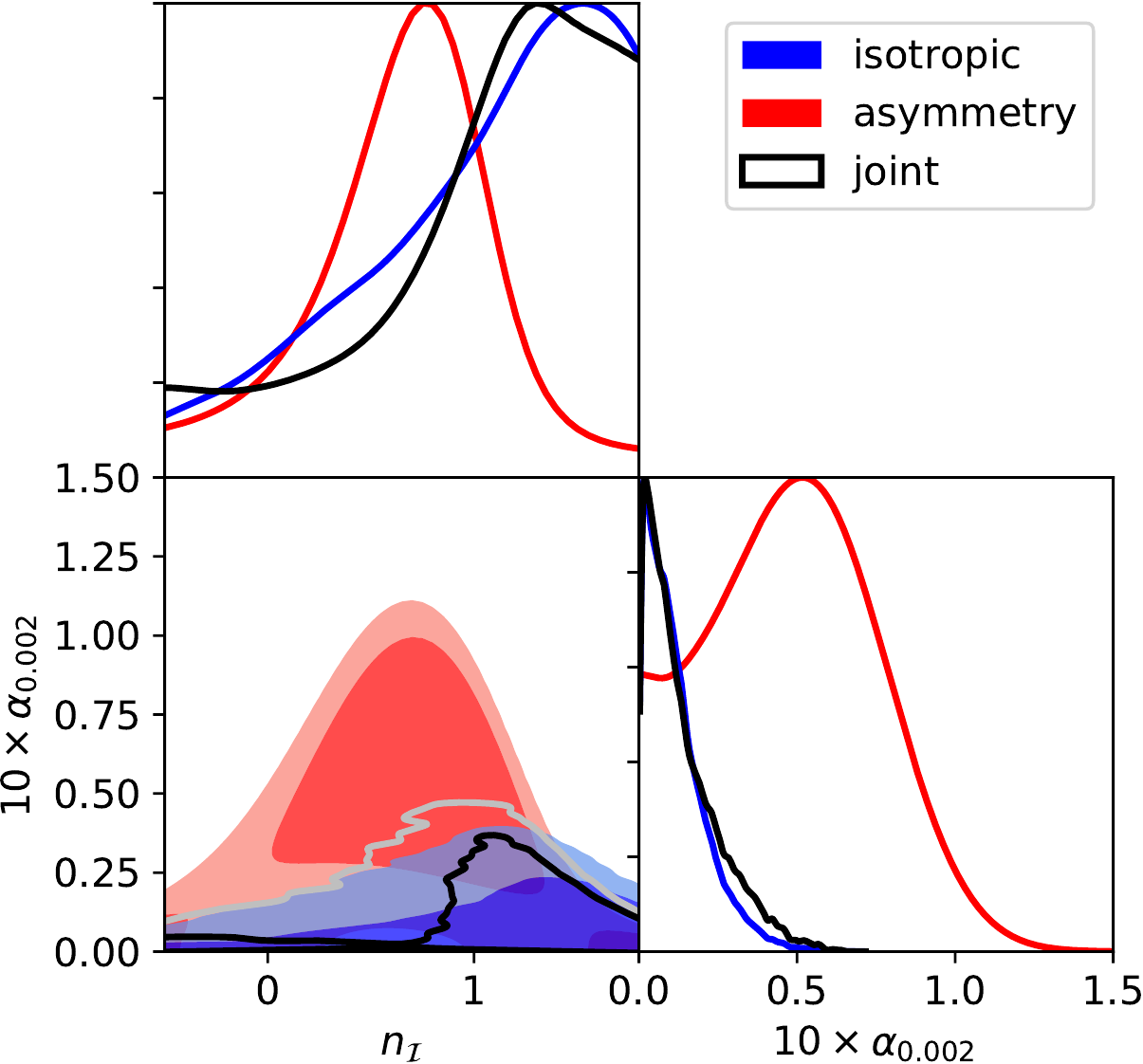}}
\centerline{\includegraphics[width=\hsize]{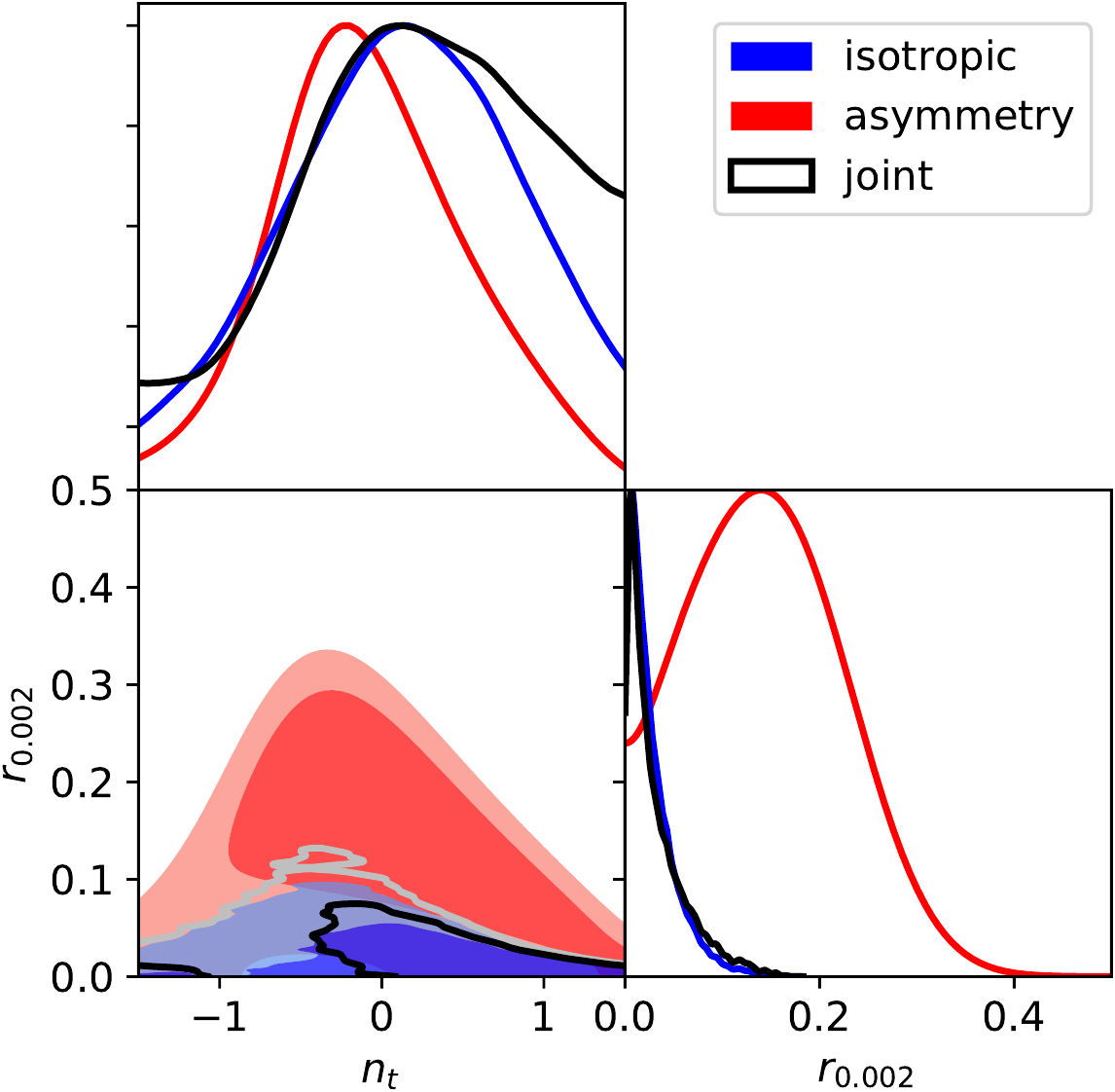}}
\caption{Posteriors for $\alpha_{0.002}$ or $r_{0.002}$ and tilt of the
isocurvature (top panel) and tensor (bottom) models.  Contours enclose $68\%$ and
$95\%$ of the posteriors.  We have conservatively assumed maximal modulation,
so that the vertical axes are also a measure of the level of
modulation relative to the isotropic $\Lambda$CDM spectrum.  We can see that the
modulation allowed by the asymmetry constraints is reduced substantially when
adding the isotropic constraints.}
\label{fig:jointconstraint}
\end{figure}

%{\em Results.---}%
   Our dipole asymmetry constraints come from \Planck\ TT data using the
\smica\ solution~\cite{planck2014-a11}.  The best-fit asymmetry spectra
for all of our models are illustrated in
\fig\ref{fig:bestfits}, where we see the expected large-scale character
of the asymmetry.  The corresponding full posteriors for $\alpha_{0.002}$
and $r_{0.002}$ and their tilts are shown in Fig.~\ref{fig:jointconstraint}
(orange contours), where we can see that large values of $\alpha_{0.002}$
or $r_{0.002}$ are needed to explain the asymmetry.  (Recall that the power
ratios $\alpha_{0.002}$ and $r_{0.002}$ also fix the modulation amplitude
for the case of maximal modulation in \eq(\ref{modflucts}).)

   For the isocurvature and tensor models we can also obtain
constraints from the isotropic power spectra described in
Table~\ref{tab:data}; we will refer to these as \emph{isotropic} constraints.
These were obtained with a version of {\tt CosmoMC}~\cite{cosmomc}
modified to accomodate uncorrelated isocurvature modes.  For these models
Fig.~\ref{fig:jointconstraint} also shows the isotropic posteriors for
$\alpha_{0.002}$ and $r_{0.002}$ and their tilts (blue contours), as well as
the joint constraints, with the assumption that the isotropic and asymmetry
likelihoods are independent (recall that they arise from diagonal
and off-diagonal elements of the multipole covariance, respectively).
Fig.~\ref{fig:jointconstraint} shows that, for both isocurvature and tensor
modulation, the joint constraints are inconsistent with the level of
modulation preferred by the asymmetry data.  In other words, the addition
of the independent isotropy data has substantially reduced the ``signal''
seen in the asymmetry data.

\begin{table}[t!]
\begin{tabular}{lccc}
\hline
\hline
Model        & Data set \\
\hline
isocurvature & \Planck\ TT,TE,EE+lowP\\
tensors      & \Planck\ TT,TE,EE+lowP+lensing+BKP\\
\hline
\end{tabular}
\caption{Data sets used for the isotropic constraints.  BKP refers to the
BICEP2/Keck Array-\Planck\ joint analysis~\cite{pb2015}.}
\label{tab:data}
\end{table}

   Note that in Fig.~\ref{fig:jointconstraint} we have assumed that the
isocurvature and tensor contributions
are maximally modulated, via \eq(\ref{modflucts}).  This allows us to directly
compare the asymmetry and isotropic posteriors, but is also a conservative
choice, because for less than full modulation the corresponding $r$ and $\alpha$
values preferred by the asymmetry constraints would necessarily be larger
with larger uncertainties.  This would increase the tension we find between
asymmetry and isotropic constraints and increase the dominance of
the isotropic data in the joint constraints.

   In order to express the above graphical results quantitatively, and
determine which models are viable for explaining the original asymmetry
signal, we will consider two quantities for each model.  The first is the
probability, $P_{>3\sigma}$, that the data allow a modulation amplitude $A$
that is at least 3 times larger than the cosmic variance $\sigma_X$.  Note that
the choice of the value 3 is arbitrary; however, if $P_{>3\sigma}$ is small
then the model cannot source significant modulation and can be ruled out,
even if $P_{>3\sigma}$ being large is an insufficient condition to prefer
a modulation model over \LCDM.
The second quantity we use is the maximum-likelihood amplitude of modulation
compared to the cosmic-variance value, $A/\sigma_X$.  For both quantities
$\sigma_X$ is calculated for asymmetry only [via Eq.~(\ref{eq:sigmax})].

   We present these quantities for the various model and data combinations in
Table~\ref{tab:probabilities}.  For the asymmetry data, both quantities are
large (except for the ISW model), which simply tells us that the models can
produce the considerable
asymmetry present in the data.  However, in all cases the values drop
substantially when adding the isotropic data.  This implies that even
maximally modulated tensor or isocurvature modes cannot source
the large asymmetry signal (or can, but with very small probability) due to
their respective isotropic constraints.
If we attempt to hide the isotropic tensor temperature power by including an
anti-correlated isocurvature mode the conclusions remain the same (see the row
marked $n_\mathrm{t} = 0^*$ in Table~\ref{tab:probabilities}).  This is due
to the different shapes of the tensor and anti-correlated isocurvature power
spectra, and not, for instance, to the nondetection of primordial $B$-modes in
the BICEP2/Keck Array-\Planck\ data.  Therefore we expect that in general,
a modulation
model for which (like the tensor and isocurvature models) isotropic power is
added will be unable to explain the dipolar asymmetry signal.
The $\tanh$, ad.-PL, and
$n_{\rm s}$-grad models are of course unaffected by the isotropic constraint
and are thus still viable modulation models as far as CMB temperature is
concerned.  For the ISW model both
$P_{>3\sigma}$ and $A/\sigma_X$ are small: even for maximal modulation the
standard \LCDM\ ISW contribution cannot explain the observed asymmetry.
Note that, via
\eq(\ref{lnLanis}), the ratio $A/\sigma_X$ is essentially the best-fit $\chi$
value, which shows that the $\tanh$ model (which has the most free parameters)
gives the best fit.

\begin{table}[h]
\begin{tabular}{lcccccc}
\hline
\hline
\multirow{2}{*}{Model}      &\multicolumn{2}{c}{Asymmetry}&\multicolumn{2}{c}{Isotropic}&\multicolumn{2}{c}{Joint}\\
                            & $P_{>3\sig}$[\%]& $A/\sig_X$& $P_{>3\sig}$[\%]&$A/\sig_X$ & $P_{>3\sig}$[\%]&$A/\sig_X$\\
\hline
tanh                        & 63.1            & 3.3       & \multicolumn{2}{c}{--}      & \multicolumn{2}{c}{--}\\
ad.-PL                      & 32.4            & 2.5       & \multicolumn{2}{c}{--}      & \multicolumn{2}{c}{--}\\
$n_{\rm s}$-grad            & 36.3            & 2.7       & \multicolumn{2}{c}{--}      & \multicolumn{2}{c}{--}\\
ISW                         & \pz0.0          & 1.2       & \multicolumn{2}{c}{--}      & \multicolumn{2}{c}{--}\\
$n_\mathcal{I}$ free        & 32.2            & 3.2       & 1.5\pz\pz       & 0.06      & 0.5\pz\pz       & 0.03\\
$n_\mathcal{I} = 1$         & 37.4            & 3.1       & 0.33\pz         & 0.10      & 1.0\pz\pz       & 0.10\\
$n_\mathcal{I} = n_{\rm s}$ & 39.6            & 3.1       & 0.073           & 0.09      & 0.24\pz         & 0.03\\
$n_{\rm t}$ free            & 29.9            & 3.1       & 0.003           & 0.03      & 0.001           & 0.02\\
$n_{\rm t} = 0$             & 37.4            & 3.1       & 0.000           & 0.48      & 0.000           & 0.63\\
$n_{\rm t} = 0^*$           & 37.4            & 3.1       & 0.000           & 0.31      & 0.000           & 0.49\\
$n_{\rm t} < 0$             & 32.1            & 3.1       & 0.008           & 0.48      & 0.003           & 0.00\\
\hline
\end{tabular}
\caption{Percentage of the posterior for which the amplitude $A$ exceeds
$3\sigma_X$, i.e., $P_{>3\sigma}$, as well as $A/\sigma_X$ for the maximum-likelihood
parameters, for different combinations of data.  These quantify whether the
model can source significant asymmetry given the data, a necessary but not
sufficient condition for preferring the model over $\Lambda$CDM.  The asterisk
denotes the addition of a fully anti-correlated isocurvature component.}
\label{tab:probabilities}
\end{table}

\begin{table}[h]
\begin{tabular}{lccc}
\hline
\hline
Model                       & Asymmetry                          & Isotropic               & Joint \\
\hline
$n_\mathcal{I}$ free        & $\alpha \leq 0.092$                & $\alpha \leq 0.031$     & $\alpha \leq 0.038$\\
$n_\mathcal{I} = 1$         & $\,0.007 \leq \alpha \leq 0.083\,$ & $\,\alpha \leq 0.038\,$ & $\,\alpha \leq 0.044\,$\\
$n_\mathcal{I} = n_{\rm s}$ & $0.008 \leq \alpha \leq 0.086$     & $\alpha \leq 0.038$     & $\alpha \leq 0.046$\\
$n_{\rm t}$ free            & $r \leq 0.28$                      & $r \leq 0.08$           & $r \leq 0.09$ \\
$n_{\rm t} = 0$             & $0.02 \leq r \leq 0.28$            & $r \leq 0.07$           & $r \leq 0.10$ \\
$n_{\rm t} = 0^*$           & $0.02 \leq r \leq 0.28$            & $r \leq 0.08$           & $r \leq 0.09$ \\
$n_{\rm t} \leq 0$          & $r \leq 0.30$                      & $r \leq 0.09$           & $r \leq 0.09$ \\
\hline
\end{tabular}
\caption{95\% CL (or upper limits) for the parameters $r_{0.002}$ and
$\alpha_{0.002}$ for various tensor and isocurvature models and data
combinations.}
\label{tab:95cl}
\end{table}

   For our best-fit parameters, the $n_{\rm s}$-grad model induces a modulation
amplitude of roughly $1.6\%$ at $k = 1\,\mathrm{Mpc}^{-1}$.  On such small
scales this model should be vulnerable to constraints from large-scale
structure surveys~\cite{gh12,fvpbmb14,bafn14,sso17,zb17}.  Indeed, this
modulation amplitude is close to (or in excess of) the 95\% upper limit based
on quasar data in~\cite{Hirata09}, and so a rigorous joint analysis may
already rule this model out.

   In order to determine quantitatively the level of modulation allowed
by the full data we look at constraints on the $r$ and $\alpha$
parameters for the isocurvature and tensor models (where we are able to use
power spectra to provide tighter constraints).  In Table~\ref{tab:95cl}
we show the 95\% CLs (or upper limits where relevant) for $r_{0.002}$ and
$\alpha_{0.002}$ for the different combinations of data.  While
the general tensor and isocurvature models (where the tilts are free to vary)
show no strong detection with the asymmetry constraints alone (in the sense
that we can only quote upper limits), we see that the addition of power spectrum data
strongly constrains the amount of modulation allowed by the data.  For models
where the tilt is fixed and not allowed to vary, the modulation signal is more
apparent; however, the addition of isotropic constraints removes the signal to
a similar degree.
Note that the asymmetry constraints in Table~\ref{tab:95cl}
allow much larger values of $r$ than $\alpha$.  This is due simply to the fact
that identical {\em primordial} ratios of tensors and isocurvature-to-adiabatic
scalar fluctuations produce much larger isocurvature
{\em temperature} fluctuations.

%----------------- CONCLUSIONS -----------------------
\section{Discussion}

%{\em Discussion.---}%
   The models we have examined fall into two general classes.  In the first,
the total statistically isotropic temperature power was constrained to match
that of standard \LCDM.  Therefore the degree of modulation could be
varied without spoiling the success of \LCDM.  In the second class,
the modulated component contributed extra power to the isotropic spectra.
Our main conclusion is that models in this latter class fail to provide
sufficient modulation to explain the dipole asymmetry without producing
too much large-scale statistically isotropic power.  Hence these models,
which include modulated tensor and uncorrelated isocurvature, can be ruled
out as the source of the large-scale dipolar asymmetry.

   Models in the first class, however, can fit the asymmetry while maintaining
the success of the \LCDM\ isotropic spectra, and hence some cannot
yet be ruled out.  One exception is a modulated ISW contribution,
which cannot source enough asymmetry to explain the signal in temperature.
The scalar tilt gradient model produces substantial modulation on small
scales, and so is at risk from survey data.  The surviving models are the
phenomenological adiabatic modulation models.  Of course
the contrived nature of such models should mean that \LCDM\ is still
preferred: they essentially add parameters to fit features in the data that
may simply be random noise.
Unfortunately a Bayesian model selection procedure would not provide
an unambiguous Bayes factor for these models, since the modulation model
evidence is strongly driven by the parameter prior ranges, which are
completely undetermined.
It will only be
possible to confirm or refute these models by comparing their predictions
for probes, (such as CMB polarization) which are sensitive to independent
fluctuation modes from CMB temperature, with future observations~\cite{long}.

%----------------- ACKNOWLEDGMENTS -----------------------
%\section*{Acknowledgments}
   This research was supported by the Canadian Space Agency, and the Natural
   Sciences and Engineering Research Council of Canada.

\bibliography{asym,Planck_bib}

\end{document}